\documentclass[angeo]{copernicus}
\usepackage{color}

\begin{document}

\title{
  A novel code for numerical 3-D MHD studies of CME expansion
}
\author[1]{Jens Kleimann}
\author[2,3]{Andreas Kopp}
\author[3]{Horst Fichtner}
\author[4]{Rainer Grauer}

\affil[1]{Nordlys Observatoriet, Universitetet i Troms\o, 9037
  Troms\o, Norway} 
\affil[2]{Institut f\"ur Experimentelle und Angewandte Physik,
  Christian-Albrechts-Universit\"at zu Kiel, 24118 Kiel, Germany}
\affil[3]{Institut f\"ur Theoretische Physik IV,
  Ruhr-Universit\"at Bochum, 44780 Bochum, Germany}
\affil[4]{Institut f\"ur Theoretische Physik I,
  Ruhr-Universit\"at Bochum, 44780 Bochum, Germany}

\runningtitle{A novel MHD code for CME expansion}
\runningauthor{J. Kleimann et al.}
\correspondence{J. Kleimann \\ (Jens.Kleimann@uit.no)}

\received{}
\pubdiscuss{} 
\revised{}
\accepted{}
\published{}


\firstpage{1}
\maketitle

\begin{abstract}
  A recent third-order, essentially non-oscillatory central scheme to
  advance the equations of single-fluid magnetohydrodynamics (MHD) in
  time has been implemented into a new numerical code. This code
  operates on a 3-D Cartesian, non-staggered grid, and is able to
  handle shock-like gradients without producing spurious
  oscillations.\\ 
  To demonstrate the suitability of our code for the simulation of
  coronal mass ejections (CMEs) and similar heliospheric transients,
  we present selected results from test cases and perform studies of
  the solar wind expansion during phases of minimum solar activity. We
  can demonstrate convergence of the system into a stable Parker-like
  steady state for both hydrodynamic and MHD winds. The model is
  subsequently applied to expansion studies of CME-like plasma
  bubbles, and their evolution is monitored until a stationary state
  similar to the initial one is achieved. 
  In spite of the model's (current) simplicity, we can confirm the
  CME's nearly self-similar evolution close to the Sun, thus
  highlighting the importance of detailed modelling especially at
  small heliospheric radii.\\ 
  Additionally, alternative methods to implement boundary
  conditions at the coronal base, as well as strategies to ensure a
  solenoidal magnetic field, are discussed and evaluated.\\
\end{abstract}

\introduction 

Coronal mass ejections (CMEs) moved into the focus of several research 
activities during recent years. Besides a variety of observational data
resulting from SOHO \citep{Pick-etal-2006}, SMEI \citep{Webb-etal-2006}, 
and, very recently, STEREO \citep{Vourlidas-etal-2007}, significant
progress has also been achieved with the numerical modelling of CMEs,
see, e.g., the reviews by \citet{Aschwanden-etal-2006} and 
\citet{Forbes-etal-2006}. 
The motivation for the various activities is at least fourfold.
First, CMEs are amongst the main mediators of the influence of the Sun on
the inner heliosphere, particularly on the Earth and its environment,
where they significantly co-determine the space weather
conditions. There is, in view of ever advancing technology that is
increasingly sensitive --- if not vulnerable --- to space weather
effects, strong interest in an understanding of the latter. An even
stronger driver of research activities is provided with the
recognition that space weather phenomena offer valuable opportunities
to study many aspects of plasma astrophysics in great detail
\citep{Scherer-etal-2005,Bothmer-Daglis-2006,Schwenn-2006}.
Second, as a consequence of the shocks driven by CMEs, they serve as
particle accelerators that do not only contribute to space weather
effects, but can be used to study the actual acceleration processes
\citep{Reames-1999, Mewaldt-etal-2005, Li-etal-2005}, which are
expected to occur in other astrophysical systems as well
\citep{Eichler-2006}.
Third, with the recent launch of the two-spacecraft mission STEREO 
\citep{Kaiser-2005}, the full three-dimensional structure of CMEs can
be observed both remotely and in-situ for the first time. First
results have already been reported by, e.g.,
\citet{Howard-Tappin-2008} and \citet{Vourlidas-etal-2007}.
And, fourth, the magnetohydrodynamic (MHD) modelling of CMEs provides
an excellent testbed for numerical codes. Although not strongly
motivated by CME physics, this is actually one of the main drivers of
model development, as is manifest with numerous approaches documented
in the literature. These various approaches can be ordered into three
groups. There is (i) principal modelling that is either analytical
and/or based on symmetry assumptions
\citep[e.g.,][]{Titov-Demoulin-1999, Roussev-etal-2003,
  Schmidt-Cargill-2003, Jacobs-etal-2005}, (ii) local modelling limited
to the extended corona, i.e.\ a few tens of solar radii
\citep[e.g.,][]{Mikic-Linker-1994}, and (iii) global modelling covering
the inner heliosphere from the solar surface out to 1~AU and beyond
\citep[e.g.,][]{Manchester-etal-2004, Odstrcil-etal-2005,
  Toth-etal-2005, Riley-etal-2006}.\\
Despite these intensified efforts and activity regarding the study of
CMEs, there remains both a number of unsolved problems and various
modelling deficiencies. For example, the acceleration and heating
processes of the plasma near the coronal base are --- even nearly 50
years after the 'discovery' of the solar wind --- still not known
\citep{Cranmer-etal-2007}, and there is also no agreement on the
processes that actually initiate CMEs \citep{Forbes-etal-2006}. Also,
their propagation and evolution in size and shape is by far not fully
understood in all detail, and neither is their interaction with the
background solar wind \citep{Jacobs-etal-2007}, with other CMEs
\citep{Gopalswamy-etal-2001}, and with planetary magnetospheres
\citep[e.g.,][]{Groth-etal-2000, Ip-Kopp-2002}. Regarding the model
formulations underlying the numerical simulation of CMEs, particularly
the (non-thermal) heating of the plasma is mostly treated in a rather
simplified manner via ad-hoc heating functions
\citep[e.g.,][]{Groth-etal-2000, Manchester-etal-2004}, variable
adiabatic indices $\gamma = \gamma({\bf r})$
\citep[e.g.,][]{Lugaz-etal-2007}, or phenomenological heating
functions \citep[e.g.,][]{Usmanov-etal-2000}, for a discussion see
\citet{Fichtner-etal-2007}.\\ 
With the intention to address several of the above-mentioned problems, 
we have applied our recently developed CWENO-based MHD code
\citep{Kleimann-etal-2004}, which primordially originated from that by 
\citet{Grauer-Marliani-2000}, to the CME expansion problem.
To our knowledge, this is the first published paper describing the
application of a CWENO-based numerical code to an MHD problem related
to space physics.\\ 
In the following, we describe the model (Sect.~\ref{sec:eqs}) and
its numerical realization (Sect.~\ref{sec:numerics} to
\ref{sec:num_boundary}), present results of analyses of both the
propagation of individual and the interaction of two CMEs, and
suggest a possible connection of our findings to observations
(Sect.~\ref{sec:runs+obs}).

\section{Governing equations}
\label{sec:eqs}

Choosing the normalization constants summarized in
Tab.~\ref{tab:norm}, the set of MHD equations for mass density
$\rho$, flow velocity ${\bf u}$, magnetic field strength ${\bf B}$,
and gas pressure $p$ reads (in dimensionless form):
\begin{eqnarray}
  \label{eq:cont}
  \partial_t \rho + \nabla \cdot (\rho \ \vec{u} ) &=& 0 \\
  \nonumber
  \partial_t (\rho \ \vec{u}) +
  \nabla \cdot \left[\rho \ \vec{u} \vec{u} \right. && \\
  \label{eq:momt} \left. + (p + \| \vec{B} \|^2/2) \
  \hat{{\mathbf I}} - \vec{B} \vec{B} \right] &=& \rho \ \vec{g} \\
  \label{eq:induct}
  \partial_t  \vec{B} + \nabla \cdot
  \left(\vec{u} \vec{B} - \vec{B} \vec{u} \right) &=& 0 \\
  \nonumber
  \partial_t e + \nabla \cdot
  \left[ (e+p+ \| \vec{B} \|^2/2) \ \vec{u} \right. && \\
    \label{eq:NRG} \left. - (\vec{u} \cdot \vec{B}) \vec{B} \right] &=&
  \rho \ (Q + \vec{u} \cdot \vec{g})
\end{eqnarray}
where
\begin{eqnarray}
  \vec{g} &=& - \Gamma/r^2 \ \vec{e}_r \quad \mbox{and} \\
  e &=&
  \frac{\rho \ \|\vec{u}\|^2}{2} +
  \frac{       \|\vec{B}\|^2}{2} +
  \left\{ \begin{array}{ccl}
    p/(\gamma-1) &:& \gamma \neq 1 \\
    0 &:& \gamma = 1
  \end{array} \right.
\end{eqnarray}
respectively denote gravity (with
\begin{equation}
  \Gamma := (G M_{\odot})/(R_{\odot} c_{\rm s}^2) = 11.49
\end{equation}
in normalized units, cf. Tab.~\ref{tab:norm}), and the total energy
density of a plasma with adiabatic exponent $\gamma$. Throughout this
paper, \mbox{$\| \cdot \|$} is used to denote the norm of a vector
(i.e.\ \mbox{$\| \vec{X} \| \equiv \sqrt{\vec{X} \cdot \vec{X}}$} for
any vector $\vec{X}$), and the symbol $\hat{\bf I}$ in
Eq.~(\ref{eq:momt}) denotes the unit tensor
(i.e.\ \mbox{$(\hat{\mathbf I})_{ij} = \delta_{ij}$}).\\ 

\begin{table}[h]
  \caption{
    \label{tab:norm} Normalization constants used for
    Eqs.~(\ref{eq:cont}--\ref{eq:NRG}). $M_{\odot}$, $R_{\odot}$,
    $k_{\rm B}$, $m_{\rm p}$, $\mu_0$, and $c_{\rm s}$ denote the
    solar mass, the solar radius, the Boltzmann constant, the proton
    rest mass, the electrical permeability of free space, and the
    isothermal sound speed, respectively.
  }
  \vskip 4mm
  \centering
  \begin{tabular}{l@{\hspace*{-12mm}}r@{\hspace*{2mm}}l}
    \tophline
    Quantity & \multicolumn{2}{c}{Normalization}\\
    \middlehline
    (solar) mass & $M_{\odot} =$ & $2.0 \times 10^{30} \ {\rm kg}$ \\
    length & $L_0 := R_{\odot} =$ & $7.0 \times 10^8 \ {\rm m}$ \\
    temperature $T$ & $T_0 :=$     & $1.0 \times 10^6 \ {\rm K} $ \\
    number density $n \equiv \rho/m_{\rm p}$  & $n_0 :=$ & $1.0
    \times 10^{14} \ {\rm m^{-3}}$ \\ 
    plasma pressure $p$
    & $2 \ n_0 \ k_{\rm B} \ T_0 =$ & $2.8 \times 10^{-3} \ {\rm Pa}$ \\
    velocity $u$
    & $c_{\rm s}:=\sqrt{2 \ k_{\rm B} \ T_0 / m_{\rm p}}=$ &
    $1.3 \times 10^5 \ {\rm m \ s^{-1}} $\\
    time $t$ & $L_0/c_{\rm s} =$   & $5.5 \times 10^3 \ {\rm s}$ \\
    energy density $e$ & $m_{\rm p} \ n_0 \ (c_{\rm s})^2 =$ &
    $2.8 \times 10^{-3} \ {\rm J \ m^{-3}}$ \\
    mag. induction $B$ & $c_{\rm s} \ \sqrt{\mu_0 \ m_{\rm p} \ n_0} =$ &
    $4.2 \times 10^{-5} \ {\rm T}$ \\
    heating rate $Q$ & $(c_{\rm s})^3 / L_0 =$ &
    $3.0 \times 10^6 \ {\rm W \ kg^{-1}}$ \\
    (CME) mass $M_{\rm cme}$ & $m_{\rm p} \ n_0 \ (L_0)^3 =$ &
    $5.7 \times 10^{13}$ kg \\
    \bottomhline
  \end{tabular}
\end{table}

A Parker-like solution for the solar wind is per construction
isothermal, i.~e.\ $\gamma=1$, resulting in an adiabatic cooling for
$\gamma>1$. In reality, the decrease in temperature $T \equiv p/ \rho$
due to this adiabatic cooling of the expanding plasma is compensated
by processes such as reconnective energy release and Alfv\'enic wave
heating. A realistic inclusion of such effects, while certainly
desirable, is beyond the scope of this first approach, and, thus,
reserved for future refinements of our model. As an alternative, we
employ an ad~hoc heating function
\begin{equation}
  \label{eq:Q_heat}
  Q = T_{\rm c} \ \alpha(r) \ (\nabla \cdot \vec{u})
\end{equation}
with a prescribed heating profile $\alpha(r)$ and a target temperature
$T_{\rm c}$. To derive the form of Eq.~(\ref{eq:Q_heat}), we first seek
the heating function $Q_{\rm iso}$ which maintains a constant
temperature $T_{\rm c}$ everywhere, irrespectively of $\gamma$. This
is done by inserting the corresponding isothermal equation of state
\begin{equation}
  p = T_{\rm c} \ \rho
\end{equation}
into the MHD equations \mbox{(\ref{eq:cont}--\ref{eq:NRG})} and
solving them analytically for $Q$, which yields 
\begin{equation}
  Q_{\rm iso} \equiv Q|_{T=T_{\rm c}}
  = (\nabla \cdot \vec{u}) \ T_{\rm c} \ .
\end{equation}
Therefore, local heating (or cooling) at any heliocentric radius
$r_{\rm h}$ is conveniently achieved by choosing $\alpha(r_{\rm h})>1$
(or \mbox{$\alpha(r_{\rm h})<1$}) in Eq.~(\ref{eq:Q_heat}). Test runs
demonstrating the validity of this method have been carried out by
\citet{Kleimann-2005}. 

\section{Numerical implementation}
\label{sec:numerics}

\subsection{Algorithm}

In order to integrate Eqs.~(\ref{eq:cont}--\ref{eq:NRG}) forward
in time, we employ a 3-D variant \citep{Kleimann-etal-2004} of a
recent semi-discrete Central weighted essentially non-oscillatory
(CWENO) scheme by \citet{Kurganov-Levy-2000} with third-order
Runge-Kutta time stepping. Notable advantages of CWENO include its
third-order accuracy in smooth regions (which automatically becomes
second order near strong gradients to minimize spurious oscillations)
and an easy generalization to multi-dimensional systems of equations
due to the fact that no (exact or approximate) Riemann solver is
needed. 
The CWENO scheme thus allows simultaneously to achieve high shock
resolution comparable with the best shock capturing schemes and high
order convergence in smooth regions dominated by plasma waves. 
Although this scheme is not strictly total variation diminishing
(TVD), simulations by \citet{Levy-Puppo-Russo-2000} do indicate an
upper bound for the total variation of their solutions. Moreover,
\citet{Havlik-Liska-2006} use a set of astrophysically relevant test
cases to compare the performance of several methods for ideal MHD and
stress CWENO's superior accuracy.  

\subsection{Example test case: Alfv\'en wings}

Various elementary tests of our implementation have been completed
successfully \citep{Kleimann-etal-2004}, such as advection in one and
two dimensions (also for propagation directions inclined at angles
\mbox{$0 < \nu < \pi/2$} to the coordinate surfaces), shock tubes with
and without magnetic field inclusion, etc. 

\begin{figure}[h]
  \vspace*{2mm}
  \begin{center}
    \includegraphics[width=8.3cm]{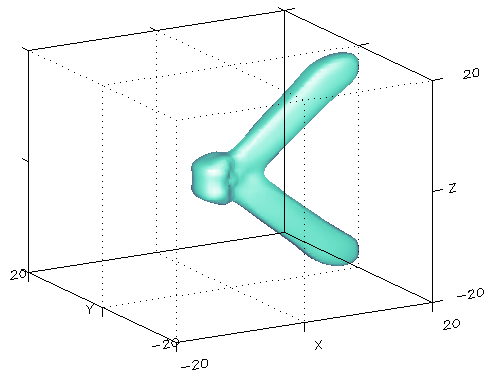}
  \end{center}
  \caption{
    \label{fig:awi-3d}
    3-D structure of a pair of Alfv\'en wings, illustrated as an
    isocontour plot of absolute velocity.
  }
\end{figure}

\begin{figure}[h]
  \vspace*{2mm}
  \begin{center}
    \includegraphics[width=0.95\hsize]{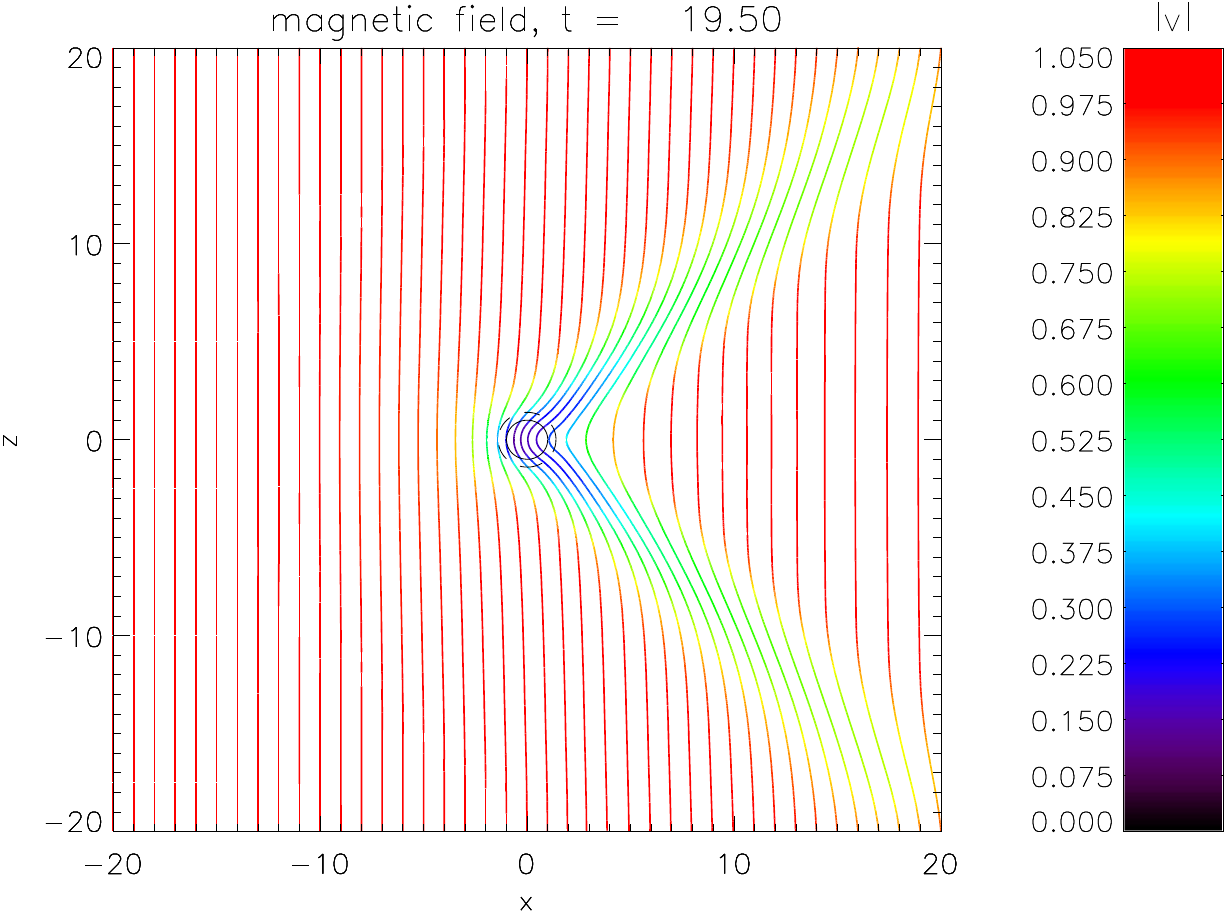}
  \end{center}
  \caption{
    \label{fig:awi-2d}
    Selected magnetic field lines in the $(\vec{u},\vec{B})$ plane,
    with color denoting flow velocity in normalized units. The flow
    is incident from the left. Since sound speed and Alfv\'en speed
    are both unity, the wings emanate in a tailward $45\degree$
    wedge when viewed in the obstacle's rest frame. The central
    circle marks the spherical volume inside of which deceleration
    according to Eq.~(\ref{eq:awings-decel}) is applied.
  }
\end{figure}

While those standard tests will not be reproduced here, one rather
advanced test setting, which is also of astrophysical relevance
 involving so-called "Alfv\'en wings" is worth being mentioned.
While the the finite extent of the wave-generating obstacle does not
  allow for an exact analytical solution, the usefulness of this
  simple but meaningful test problem stems from the fact that it
  incorporates several types of MHD waves (Alfv\'en and slow/ fast
  magnetosonic), the expansion speed and characteristics of which can
  be verified quantitatively with theoretical expectations to ensure
  proper implementation of the relevant physics.  
As shown by \citet{Drell-etal-1965}, the movement of a conductive
obstacle (e.g. a satellite or small planet) through a homogeneous
fluid with a perpendicular magnetic field will generate standing MHD
waves in the $(\vec{u}, \vec{B})$ plane called Alfv\'en wings. This
phenomenon plays a major role for the interaction of the moon Io with
the Jovian magnetosphere \citep{Neubauer-1980,Linker-etal-1988}, and
also for artificial satellites in the magnetosphere of Earth, see
e.~g.\ \citet{Kopp-Schroeer-1998} and references therein.\\ 
The corresponding numerical test case, which works well both in
  2-D and 3-D, involves an initially
homogeneous flow \mbox{$\vec{u}=u_0 \ \vec{e}_x$} of constant density,
which is combined with a perpendicular, equally homogeneous magnetic
field \mbox{${\bf B}=B_0 \ {\bf e}_z$}. A solid, spherical obstacle is
then implemented by performing an artificial deceleration 
\begin{equation}
  \begin{array}{rl}
    \label{eq:awings-decel}
    \vec{u}(\vec{r},t) \leftarrow \vec{u}(\vec{r},t)
    & \times \left[ 1-{\rm min}(t,1) \right] \\
    & \times \left[ 1-\tanh (4 \ {\rm max} (\| \vec{r} \|-1,0 )) \right]
  \end{array}
\end{equation}
after each time step, such that for $t \ge 1$, the flow will vanish
within \mbox{$\|\vec{r}\| \le 1$}. Figures~\ref{fig:awi-3d} and
\ref{fig:awi-2d} illustrate the emanating wing structure. Direction
and speed of propagation agree well with their respective theoretical
expectations.

\subsection{Choice of coordinates}

At first sight, the Sun's obviously spherical shape would suggest the
use of spherical coordinates \mbox{$[r,\vartheta,\varphi]$},
especially since the radial convergence of lines of constant
$\vartheta,\varphi$ entails the additional benefit of increased
spatial resolution near the Sun's surface. On the other hand, the
Courant-Friedrichs-Lewy (CFL) criterion of numerical validity and
stability \citep{Courant-etal-1928}, which requires the ``velocity''
\mbox{$\Delta x / \Delta t$} to be greater than the maximum physical
propagation velocity, imposes a limit on the time step $\Delta t$
based on the cell size $\Delta x$.
The very choice of a coordinate system with varying grid cell sizes,
together with the requirement that the time step be uniform on the
entire grid, thus implies that $\Delta t$ will be set by the
\mbox{$\Delta x$} of the grid's smallest cell. For spherical
coordinates, this means that the increased resolution at small $r$,
however welcomed for physical reasons, would force \mbox{$\Delta t$}
to be much lower than what the CFL criterion would require for most
parts of the computational domain. This 'problem of small time steps'
is avoided by using Cartesian coordinates, which have equal cell
spacing everywhere and thus do not waste computing time on the larger
cells. Even worse is the problem of coordinate singularities at the
poles \mbox{$\vartheta \in \{0,\pi \}$}, which require delicate
numerical treatment. For these reasons, we opt for Cartesian
coordinates \mbox{$[x,y,z]$}, for which numerics are faster, simpler
(esp.\ with respect to multi-dimensional extension), and more
stable. This is especially true since our CWENO code is built within a
framework that allows for Cartesian Adaptive Mesh Refinement (AMR, see
\citep{Kleimann-etal-2004}). This is of high interest for more
detailed studies of, e.g., the inner structure of a CME.
While AMR can, in principle, be used with spherical coordinates, its
advantage is over-compensated by the fact that the convergence of grid
spacing implies unacceptably low CFL numbers. (Note also that since
CMEs generally do not exhibit any clear spatial symmetry, the use of
non-Cartesian coordinates is not expected to entail any particular
advantages for their description.) 

\subsection{Divergence cleaning}

Like many other algorithms, CWENO does not exactly conserve the
solenoidality condition \mbox{$\nabla \cdot {\bf B}=0$} for the
magnetic field, and a correction scheme becomes mandatory to avoid
unphysical artifacts. From the wealth of existing schemes (for an
overview see, e.~g., \citet{Toth-2000}), we have evaluated the
performance of the Generalized Lagrange Multiplier (GLM) approach by
\citet{Dedner-etal-2002} against a classical projection scheme (see
Sect.~\ref{sec:proj}).\\

\subsubsection{The GLM scheme}

\begin{figure*}[t]
  \begin{center}
    \includegraphics[width=17.5cm]{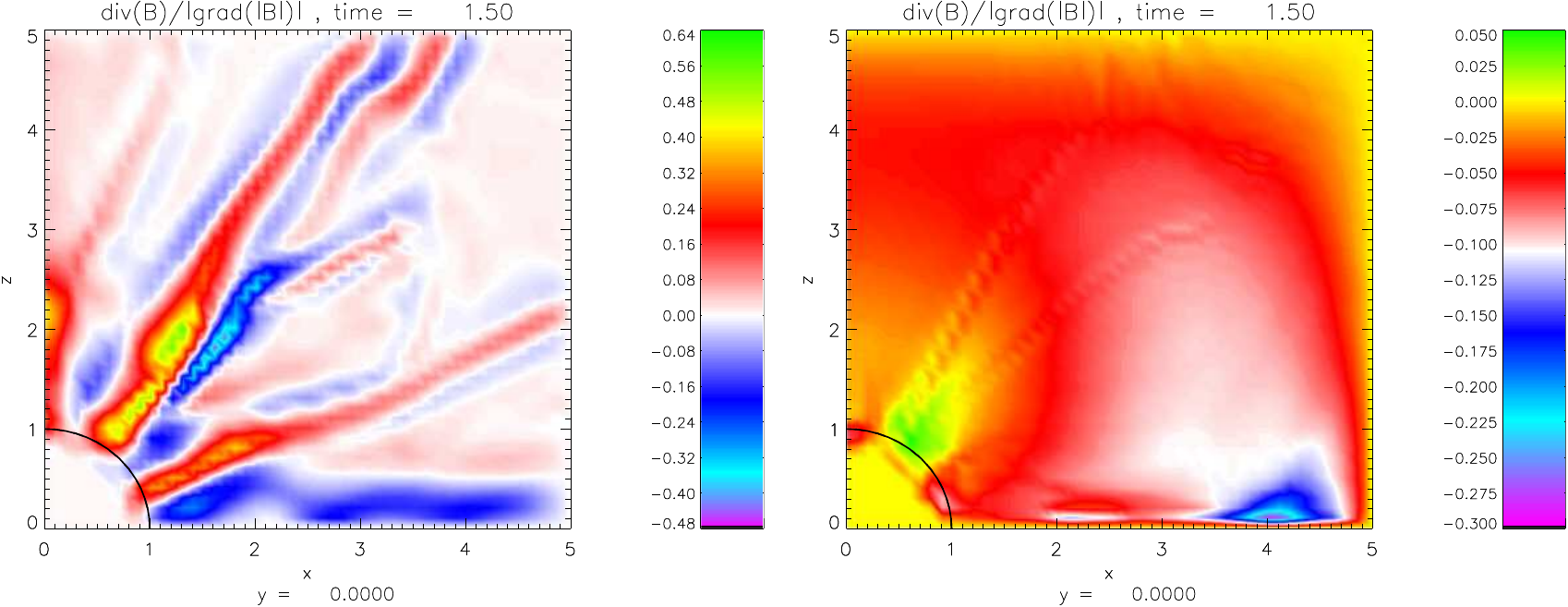}
  \end{center}
  \caption{
    \label{fig:divB}
    Normalized divergence error
    \mbox{$\kappa := (\nabla \cdot \vec{B}) / \|
      \nabla \sqrt{\vec{B} \cdot \vec{B} } \|$} in the (poloidal)
    $(x,z)$ plane for a standard solar wind run at time $t=1.5$
    without correction (left) and using GLM (right). The improvement
    is substantial but still insufficient due to massive divergence
    values introduced at the inner boundary (unit circle around the
    origin). The projection scheme achieves $\kappa \sim 10^{-6}$ (not
    shown). Note the different color scales.
  }
\end{figure*}

The GLM scheme solves an additional equation
\begin{equation}
  \label{eq:psi_clean}
  \partial_t \Psi + (v_{\rm f})^2 \ \nabla \cdot \vec{B} = -
  ( v_{\rm f} / \lambda ) \ \Psi
\end{equation}
for a position- and time-dependent Lagrange multiplier $\Psi$, and
adds a term $-\nabla \Psi$ to the right hand side of
Eq.~(\ref{eq:induct}). This procedure causes $\Psi$ (and hence
\mbox{$\nabla \cdot \vec{B}$}) to be damped with decay constant
 \mbox{$\tau_{\rm d}:= \lambda / v_{\rm f}$}, while at the same time
 advection of $\Psi$ towards the boundary of the computational volume
 occurs at the highest permissible speed $v_{\rm f}$ (chosen to equal
 the global maximum of the fast magnetosonic speed in this
 case). Following \citet{Dedner-etal-2002}, a value of 0.18 is used
 for the second constant $\lambda$.\\  
The main advantage of this method is that Eq.~(\ref{eq:psi_clean})
already possesses the correct conservative form, allowing for direct
treatment with CWENO. In particular, physical conservation laws are
not affected in any way. \\
Figure~\ref{fig:divB} compares the performance of the two methods for
a standard run. The obviously inferior performance of GLM can be
explained by the fact that within a spherical layer ${\cal L}$ around
the inner (solar) boundary, the boundary procedure described in
section~\ref{sec:inner_b} entails an averaging of the inner boundary
value $\vec{B}_{\rm in}$ and the newly computed outer solution
$\vec{B}_{\rm out}$ via 
\begin{equation}
  \vec{B}_{\rm avg} := f \vec{B}_{\rm in} + (1-f) \vec{B}_{\rm out}
\end{equation}
for some function $f:{\cal L} \mapsto [0,1]$, which is bound to
introduce a marked violation of the divergence constraint due to the
first term of
\begin{equation}
  \nabla \cdot \vec{B}_{\rm avg} = \nabla f \cdot \left( \vec{B}_{\rm in} -
  \vec{B}_{\rm out} \right) - f \ \nabla \cdot \vec{B}_{\rm out}
\end{equation}
being clearly non-zero. This divergence-laden field is then advected
outwards by the wind flow, thus causing the magnetic field to quickly
become non-solenoidal in the outer region as well. (This behavior
becomes particularly evident in the left plot of
Fig.~\ref{fig:divB}.)\\  
Since the resulting magnitude of \mbox{$\nabla \cdot \vec{B}$} in the
non-solenoidal interface layer is inversely proportional to the
layer's thickness, the problem cannot be avoided by choosing a
different matching method (i.~e.\ a different matching function
\mbox{$r \mapsto f(r)$}). Note that this line of reasoning includes
the case of doing no averaging at all: This simply corresponds to the
limiting case $f=f_{\rm step}$, where 
\begin{equation}
  f_{\rm step}: \vec{r} \mapsto \left\{
    \begin{array}{rcl}
      1 &:& \|\vec{r}\| \le 1 \\
      0 &:& \|\vec{r}\| > 1
    \end{array}
  \right. \ .
\end{equation}
Since this non-solenoidal layer is actively re-created every time the
newly computed outer solution is connected to the inner boundary, we
may conclude that a suitable divergence cleaning procedure must kill
the divergence immediately afterwards in one step (as the projection
scheme does), rather than only damping/ transporting it away on
somewhat longer timescales (GLM). \\
We must therefore conclude that for investigations of this kind, the
presence of an inner inflow boundary is, at least, difficult and may, in
some cases, even preclude the use of the GLM scheme for divergence
cleaning. (Note however, that the applicability of GLM to other
settings lacking such an internal boundary remains unimpeded by this
finding.) \\

\subsubsection{The projection scheme}
\label{sec:proj}

The so-called 'projection method' was originally developed by
\citet{Chorin-1967} for simulations of inviscid flow, and later
applied in the context of MHD simulations by
\citet{Brackbill-Barnes-1980}. It solves the Poisson equation
\begin{equation}
  \nabla^2 \Phi = \nabla \cdot \vec{B}
\end{equation}
for $\Phi$ and then subtracts $\nabla \Phi$ from $\vec{B}$ to ensure
\mbox{$\nabla \cdot \vec{B}=0$}. While numerically expensive, it is
able to reduce divergence errors down to machine accuracy, and will
therefore be used in all simulations presented here.

\section{Boundary and initial conditions}
\label{sec:boundary}

\subsection{Types of boundaries}

The computational volume consists of a brick-shaped region of space
covering \mbox{$ 100 \times 70 \times 50$} cells in the $x$, $y$, and
$z$ direction, respectively. Each cell is a cube with a side length of
typically \mbox{$\Delta x = \Delta y = \Delta z = 0.1$}, implying a
coverage of $[5,7,10] \ R_{\odot}$ of real space. (We note that this
relatively coarse resolution was chosen deliberately do demonstrate
the excellent symmetry-maintaining properties of the employed scheme,
see also Fig.~\ref{fig:vpark} of Sect.~\ref{sec:extrapol_v}. Higher
spatial resolution, however desirable for the study of fine-scale
structures, would tend to diminish the magnitude of numerical
artifacts by which the scheme's performance could be judged, thereby
hampering the usefulness of this demonstration.)   
The Sun's center is located at the origin, with the dipolar axis
pointing into the positive $z$ direction. The computational domain is
surrounded by two layers of 'ghost cells', whose values are updated
after each time step either from symmetry considerations (for 'mirror'
boundaries intersecting the origin), or use of outflowing boundary
conditions (at the actual 'outer' boundaries).\\
The solar surface, which is represented by a sphere of unit radius
located inside the computational volume, obviously does not coincide
with any of the Cartesian coordinate surfaces, and therefore requires
special treatment, which is discussed in detail in
Sect.~\ref{sec:inner_b} and \ref{sec:extrapol_v}.
This inner boundary is particularly delicate since it constitutes the
surface from which the solar wind emanates, such that numerical
artifacts imposed by an imperfect treatment of this boundary will be
quickly advected through the entire domain. 

\subsection{Initial conditions}
\label{sec:init}

The generic setup for quiet-Sun solar wind simulations is as follows:
At $t=0$, the simulation is initialized with a radially symmetric wind
flow \mbox{$\vec{u}(\vec{r})=u(r) \ \vec{e}_r$} with
\begin{equation}
  \label{eq:v_start}
  u(r) = \frac{u_{\rm m}}{2 r_{\rm m}-3} \times \left\{
  \begin{array}{ccl}
    0 &:& r < 1 \\
    (r-1)^2 &:& 1 \leq r \leq 2 \\
    2 r-3   &:& r > 2
  \end{array} \right.
\end{equation}
such that a super-sonic value ${u_{\rm m}}$ is reached at the
innermost boundary point, $r_{\rm m}$. This is done to ensure that the
initial velocity at the outer boundary is as small as possible to
allow large time steps \mbox{$\Delta t$}, while at the same time being
large enough to prevent numerical boundary artifacts from being
transported inwards.\\ 
The density scales as $\rho(r) \propto 1/r^3$, and the temperature is
equal to a constant $T_c$. The initial magnetic field is implemented
using
\begin{equation}
  \label{b-init}
  \vec{B}|_{t=0} = \nabla \times
  \left( \frac{F(r)}{r} \sin \vartheta \ \vec{e}_{\varphi} \right)
\end{equation}
where $F(r)=P_0/r$ yields a dipole of strength $P_0$ that is aligned
with the $z$~axis.\\
Since the projection scheme described in Sect.~\ref{sec:proj} will
operate on the entire grid, the singularity of Eq.~(\ref{b-init})
at \mbox{$r=0$} must be avoided. This is achieved by choosing a
suitably matched polynomial for $F(r)$ inside some small sphere around
the origin. (Note that the radius of this sphere must be chosen at
least several grid cell sizes smaller than unity to prevent the
non-zero current density associated with \mbox{$F(r) \ne P_0/r$} from
causing unphysical Lorentz force accelerations just outside the $r=1$
boundary.)

\section{Numerical treatment of the solar surface boundary}
\label{sec:num_boundary}

\subsection{The interpolation method}
\label{sec:inner_b}
  
\begin{figure}
  \vspace*{2mm}
  \begin{center}
    \includegraphics[width=8.3cm]{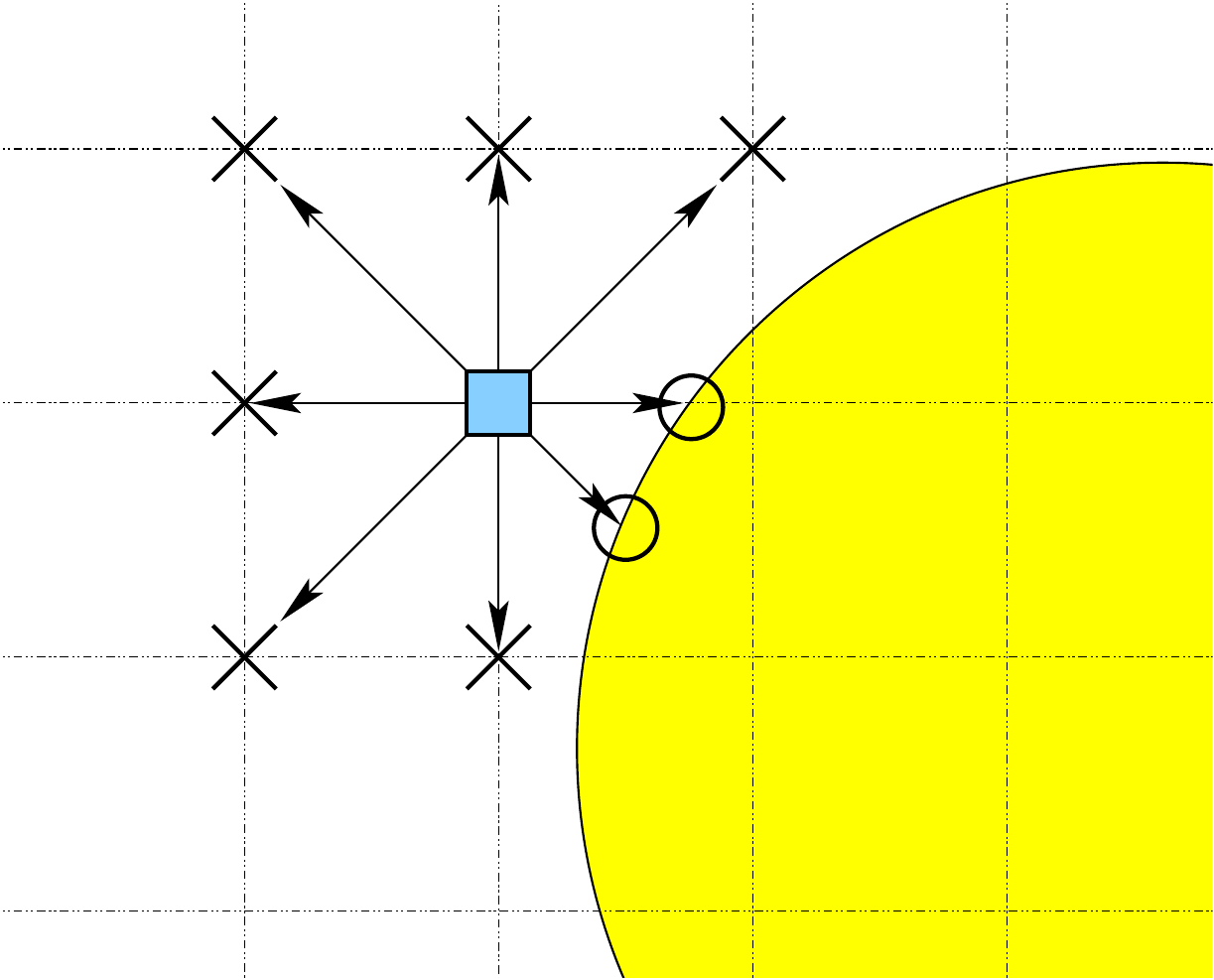}
  \end{center}
  \caption{
    \label{fig:forcing}
    2-D analog of the averaging procedure. At each grid point
    \mbox{$\vec{r}_I \in {\cal I}$} (shaded central box), a weighted
    average is computed from time-advanced values taken at neighbors
    outside ${\cal S}$ (crosses), and boundary values taken in the
    direction of neighbors inside ${\cal S}$ (circles). The factors
    $L_{I \alpha}$ entering into Eq.~(\ref{eq:avg}) are equivalent to
    the length of arrows in the diagram. 
  }
\end{figure}

The inner (solar surface) boundary, which is just the sphere
\mbox{${\cal S}:=\{ \vec{r}| \ \|\vec{r}\|=1 \}$}, obviously does not
coincide with any of the Cartesian coordinate planes, which brings up
the question of how these boundary conditions are best represented on
the grid.
Simple-minded attempts, such as keeping cell values inside the Sun
fixed and integrating only those outside, have been tried but were
shown to result in block-like artifacts at small radii (essentially
tracing the envelope of the set of grid cells considered 'inside')
where the problem's symmetry would stipulate spherical contours. While
these artifacts would of course diminish as spatial resolution is
increased, it seems vital to obtain a high degree of symmetry-keeping
already at this relatively coarse resolution, especially in view of
the high numerical costs associated with increasing the number of grid
cells in a 3-D simulation.\\ 
After several possibilities have been tried, the following procedure
was adopted:  
\begin{enumerate}
\item At initialization, all grid points which are located outside
  ${\cal S}$ but have at least one of their $3^3-1=26$ neighbors
  inside ${\cal S}$ are stored in a list ${\cal I}$ of 'interface
  points'. (The set neighbors of a cell $\vec{r}_{ijk}$ is defined as
  the set of cells \mbox{$\vec{r}_{i^{\prime} j^{\prime} k^{\prime}}$}
  with
  \mbox{$|i-i^{\prime}|,|j-j^{\prime}|,|k-k^{\prime}| \in \{0, 1 \}$} 
  excluding $\vec{r}_{ijk}$ itself.)
\item After each time step (which only advances grid points outside
  ${\cal S}$ in time), a weighted average for each variable 
  \mbox{$w \in \{\rho, \rho u_x, \rho u_y, \rho u_z, B_x, B_y, B_z, e \}$}
  is computed for each $\vec{r}_I \in {\cal I}$ via
  \begin{equation}
    \label{eq:avg}
    \bar{w}_I = \left( \sum_{\alpha}
    (L_{I \alpha})^{-1} \ w(\vec{r}_{\alpha}^{\prime}) \right)
    \left/ \left( \sum_{\alpha} (L_{I \alpha})^{-1} \right) \right.
  \end{equation}
  with
  \begin{equation}
    \vec{r}_{\alpha}^{\prime} = \left\{
    \begin{array}{ccc}
      \vec{r}_{\alpha}
      &:&
      \mbox{$r_{\alpha}$ outside of ${\cal S}$} \\
      \overline{\vec{r}_I \ \vec{r}_{\alpha}} \ \cap \ {\cal S}
      &:&
      \mbox{$\!\!\!r_{\alpha}$ inside of ${\cal S}$}
    \end{array} \right.
  \end{equation}
  and
  \mbox{$L_{I \alpha}:= \|\vec{r}_I-\vec{r}_{\alpha}^{\prime} \|$},
  where the sums in Eq.~(\ref{eq:avg}) are taken over all
  neighbors of $\vec{r}_I$. The choice of weights
  \mbox{$\propto (L_{I \alpha})^{-1}$} ensures that for
  \mbox{$ \|{\bf r}_I \| \rightarrow 1$}, $\bar{w}_I$ smoothly tends
  to the appropriate boundary value. Figure~\ref{fig:forcing} serves
  to illustrate the situation.\\
  When the above procedure is applied to the Cartesian components of
  vectors such as $\vec{u}$  and $\vec{B}$, it will usually destroy
  any possibly existing symmetry of these vector fields (e.~g., if
  $\vec{u}$ is purely radial, the averaged $\bar{\vec{u}}$ vectors
  will slightly deviate from the radial direction). In order to
  preserve such symmetries, all Cartesian vector components entering
  the averaging process of Eq.~(\ref{eq:avg}) are first rotated
  until they are parallel to $\vec{r}_I$ before the averaging takes
  place, thus ensuring that the symmetry is preserved. 
\item In order to guarantee that the newly computed grid values for
  ${\cal I}$ are independent of the ordering within that list, all
  computed averages are first stored in a separate field. Only when
  all the $\bar{w}_I$ are known will they be copied onto the actual
  grid.
\end{enumerate}
Note that step 1 is executed only once, while steps 2 and 3 are called
after each integration time step.\\ 
The above procedure gives the best results when applied to a scalar
field that varies approximately linear in space. Near the solar
surface, however, strong radial gradients of density are present. For
this reason, it has been found to be advantageous to artificially
reduce the density gradient in Eq.~(\ref{eq:avg}) by multiplying
$\rho(\vec{r}_{\alpha}^{\prime})$ with 
$\| \vec{r}_{\alpha}^{\prime} \|^{3...4}$ before averaging, and
consequently dividing $\bar{\rho_I}$ by $\| \vec{r}_I \|^{3...4}$
afterwards.

\subsection{Velocity extrapolation versus fixed boundary}
\label{sec:extrapol_v}

\begin{figure*}
  \vspace*{2mm}
  \begin{center}
    \includegraphics[width=17.5cm,height=8cm]{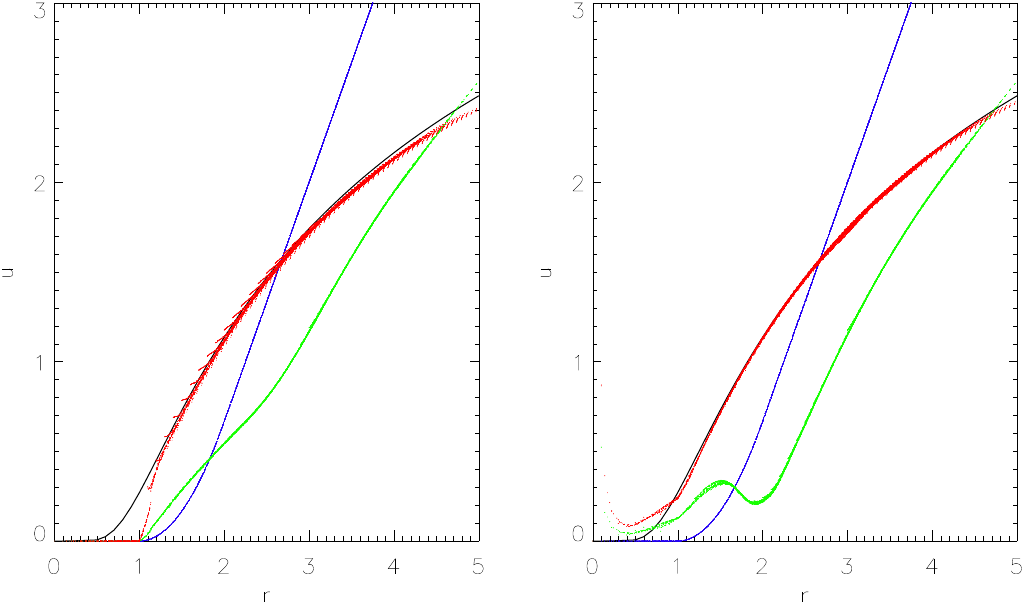}
  \end{center}
  \caption{
    \label{fig:vpark}
    Comparison of two possible methods to impose a boundary condition
    for $\vec{u}$ at the inner boundary $r=1$: Dirichlet boundary
    condition (left) enforcing \mbox{$\vec{u}|_{r=1} = \vec{0}$}
    versus inward extrapolation (right). Both diagrams show scatter
    plots of absolute velocity $\| \vec{u} \|$ versus heliocentric
    radius for all $40^3$ grid points as the system converges towards
    the isothermal Parker wind solution (black solid curve). Colors
    are used to denote the moment of initialization ($t=0$, blue), an
    intermediate step ($t=1$, green), and the situation after the
    stationary equilibrium has been reached ($t=8$, red). Since all
    grid points are shown, the scatter at a given radius can be seen
    as a measure of the simulation's spurious departure from radial
    symmetry. 
  }
\end{figure*}

The averaging procedure of Sect.~\ref{sec:inner_b} keeps all
quantities fixed on ${\cal S}$. However, if solar wind configurations
such as the Parker wind solution \citep{Parker-1958} are to be
reproduced, it seems questionable to apply this procedure to the
velocity, since the requirement that \mbox{$r \mapsto \|\vec{u}(r)
  \|$}  must pass through a critical (sonic) point completely
determines the solution topology, and thus eliminates the freedom to
prescribe a fixed (Dirichlet) boundary value at $r=1$.\\ 
Different possibilities are conceivable to handle this problem:
\begin{enumerate}
\item Allow the velocity near ${\cal S}$ to adjust freely by radial
  inward extrapolation of the time-advanced solution
  \citep{Keppens-Goedbloed-2000}, or
\item enforce a fixed value for $\vec{u}$ on ${\cal S}$ in spite of
  the above problem, and accepting that (hopefully small) inaccuracies
  will be introduced at small radii \citep{Manchester-etal-2004}.
\end{enumerate}
While the second alternative is just what the above averaging
procedure does, the first option, while being straightforward in
spherical coordinates, is clearly non-trivial to implement on the
present Cartesian grid.\\ 
In analogy to the averaging scheme used for the other variables, the
adopted procedure (which replaces the procedure of
Sect.~\ref{sec:inner_b} for $\vec{u}$) is as follows:
\begin{enumerate}
\item Prior to initialization, a list ${\cal A}$ of all grid points
  $\vec{r}_{A}$ with \mbox{$1 \le \| \vec{r}_{A} \| \le 0.5$} is set
  up and sorted by decreasing $\| \vec{r}_{A} \|$ (such that the
  outermost points will be processed first).
\item For each $\vec{r}_{A} \in {\cal A}$, a sub-list of grid points
  $\vec{r}_{A,i}$ is created, such that
  \begin{itemize}
  \item[(i)]  $\| \vec{r}_A \| < \| \vec{r}_{A,i} \|$ \quad and
  \item[(ii)] $\| \vec{r}_A - \vec{r}_{A,i} \| < r_0$ \quad
    (with \mbox{$r_0 \approx (2...3) \ \Delta x$)}.
  \end{itemize}
  In other words, the sub-list for $\vec{r}_A$ contains grid points
  close to $\vec{r}_{A}$ which are located at larger radii than
  $\vec{r}_{A}$ itself. (Note again that steps 1 and 2 are executed
  only once.)
\item After each time step, the radial mass flux
  \begin{equation}
    f_{A,i} := (\rho \vec{u})_{A,i} \cdot
    \vec{r}_{A,i} \ \| \vec{r}_{A,i} \|
  \end{equation}
  is computed from the sub-list at each $\vec{r}_{A}$, and a
  least-squares fit of the linear function
  \mbox{ $g_A: r \mapsto c_{0,A}+c_{1,A} \ r$} is used to find the
  mass flux at $\vec{r}_A$ (which is then given by
  \mbox{$g_A(\|\vec{r}_A\|)=c_{0,A}+c_{1,A} \|\vec{r}_A\|$}). Finally,
  the corresponding radial momentum is immediately afterwards written
  to the grid, such that its value is available to the extrapolation
  at the next point in the list.
\end{enumerate}
Figure~\ref{fig:vpark} shows a comparison of both methods for the
(unmagnetized) Parker wind case. The interpolation method's superior
performance is in the range of a few percent only and has to be gauged
against its increased computational effort. Consequently, all of the
simulations presented here use the Dirichlet method. (We note,
however, that this may not always be appropriate when different
parameter ranges are used. For instance, a higher base temperature
$T_c$ will move the sonic point sunwards, leading to a higher flow
velocity at the solar surface, and a presumably larger discrepancy to
the zero-velocity condition.) 

\section{Solar wind and CME simulations}
\label{sec:runs+obs}

\subsection{Creating equilibrium wind solutions}

For the initialization of our CME expansion studies, we first seek a
well-defined MHD equilibrium resembling a 'quiet' (i.e.\ stationary)
setting during solar minimum. 
While this is of course not strictly required for such studies, it is
nevertheless vital for the interpretation of the obtained results,
since only then can structures like CMEs be clearly disentangled from
the dynamics of the background flow.\\ 
For magnetized, isothermal ($\gamma=1$) winds, the system starts from
the initial conditions of Sect.~\ref{sec:init} and then quickly
(within a few sound crossing times) settles into a stable equilibrium
similar to the one depicted in the first frame of
Fig.~\ref{fig:cme-pixpanel}. At a distance of 5~$R_{\odot}$ from the
origin, the outflow velocity in $x$ direction differs from that at the
polar field line by a factor of about
  \[
    \frac{\|{\bf u}(5,0,0)\|}{\|{\bf u}(0,0,5)\|} =
    \frac{1.26 \ c_{\rm s}}{3.13 \ c_{\rm s}} \approx
    \frac{160 \ \mbox{km/s}}{400 \ \mbox{km/s}} = 0.4 \ ,
  \]
which is due to the retaining force of the closed magnetic field lines
in the equatorial region, and reminiscent of the speed difference
between the fast and slow solar wind. Examples of non-isothermal
hydrodynamical runs integrating the full energy equation
(\ref{eq:NRG}) with the heating source term (\ref{eq:Q_heat}) can be
found in \citep{Kleimann-2005}.\\ 
It is noteworthy that essential features of the quiet inner
heliosphere, such as a latitudinal dependence of outflow velocities
resembling the fast and slow solar wind and the poleward transition
from closed magnetic field lines (which span a static 'dead zone')
below about $40\degree$ of latitude to an open, more radial field, are
self-consistently reproduced by our model. In particular, it was found
to be unnecessary to invoke the method of latitude-dependent inner
boundary conditions used by other authors
\citep{Keppens-Goedbloed-2000, Manchester-etal-2004} to reproduce this
dichotomy: The magnetic dipole strength $P_0$ proved fully sufficient
to control the latitudinal extent of the closed-field helmet zone. As
can be intuitively expected, a stronger B field at the surface will
tend to conserve its arch-shaped closed structure, while in the limit
\mbox{$P_0 \rightarrow 0$}, all field lines will be stretched out
radially by the flow, and spherical symmetry is recovered. The choice
of $P_0=4$ results in the intermediate case with an open/ closed
transition near $\pm 40\degree$ of solar latitude.

\subsection{Initialization of CME onset}

The present investigation focuses on the aspect of CME propagation,
rather than on their actual nascency. Therefore, a simplifying
approach similar to the one already employed by
\citet{Groth-etal-2000} and \citet{Keppens-Goedbloed-2000} will be
used. This approach is based on a time-dependent boundary condition at
the solar surface, generating a transient, isothermal increase in
density (and thus in pressure). If chosen sufficiently strong, this
density excess is able to tear open the equatorial helmet streamer,
causing the detachment of the excess matter as a rapidly expanding
bubble.\\  
In order to initiate an eruption in the time interval
\begin{equation}
  {\cal T} := [t_{\rm cme}, t_{\rm cme}+\tau_{\rm cme}]
\end{equation}
an additional, localized mass flux
\mbox{$\rho_{\rm add} \ \vec{u}_{\rm add}$} with 
\begin{equation}
  \begin{array}{rcl}
    \rho_{\rm add}    (\vec{r},t)|_{r=1} &=& \rho_{\rm cme}(\vec{r},t) \\ 
    \vec{u}_{\rm add} (\vec{r},t)|_{r=1} &=&    u_{\rm cme} \ \vec{e}_r
  \end{array}
\end{equation}
is released at a pre-defined location on the solar surface (implying
\mbox{$\|\vec{r}\|=1$} for the remainder of this section). Without
loss of generality, let the center of the eruption region be in the
plane $\varphi=0$, such that its location is just 
\begin{equation}
  \vec{r}_{\rm cme} :=
  \left(\begin{array}{c}
    x_{\rm cme} \\ y_{\rm cme} \\ z_{\rm cme}
  \end{array} \right) =
  \left(\begin{array}{c}
    \sin \vartheta_{\rm cme} \\ 0 \\ \cos \vartheta_{\rm cme}
  \end{array} \right) \ .
\end{equation}
For fixed time $t$, the value of $\rho_{\rm cme} (\vec{r},t)$ should
only depend on the angular distance
\begin{equation}
  \alpha(\vec{r},\vec{r}_{\rm cme})
  = \arccos (\vec{r} \cdot \vec{r}_{\rm cme})
\end{equation}
between $\vec{r}$ and $\vec{r}_{\rm cme}$, such that
$\rho_{\rm cme}(\vec{r},t)$ possesses axial symmetry with respect to
the $\vec{r}_{\rm cme}$~axis. Following
\citet{Keppens-Goedbloed-2000}, we employ the function 
\begin{equation}
  \label{eq:density_ex}
  \rho_{\rm cme}(\vec{r},t) := \left\{
  \begin{array}{ccl} \displaystyle
    f_0 \ E(\vec{r},t) &:&
    t \in {\cal T} \wedge \alpha \le \delta_{\rm cme} \\ && \\
    0 &:& \mbox{else}
  \end{array} \right.
\end{equation}
with
\begin{equation}
  \label{eq:density_ex_D}
  E(\vec{r},t) :=
  \sin^2 \left(\pi \frac{t-t_{\rm cme}}{\tau_{\rm cme}} \right) \
  \cos^2 \left(\frac{\pi}{2} 
  \frac{\alpha(\vec{r},\vec{r}_{\rm cme})}{\delta_{\rm cme}} \right)
\end{equation}
which connects smoothly to the undisturbed state in both space and
time. Here $2 \delta_{\rm cme}$ denotes the angular diameter of the
circular eruption region $\Omega_{\rm cme}$, which is defined as the
region where $\rho_{\rm cme}({\bf r},t)$ gives a non-zero contribution
according to Eq.~(\ref{eq:density_ex}), and which thus covers a total
solid angle  
\begin{equation}
  \omega_{\rm cme} := \int\limits_0^{\delta_{\rm cme}}
  2\pi \sin \alpha \ {\rm d} \alpha 
  = 2\pi \left[1-\cos (\delta_{\rm cme}) \right]
\end{equation}
on the Sun's surface. The total mass released by the CME's eruption
can be estimated as 
\begin{eqnarray}
  M_{\rm cme} &:=& 
  \int\limits_{\cal T} \int\limits_{\Omega_{\rm cme}}
  \rho_{\rm cme}(\vec{r},t) \ u_{\rm cme} \ {\rm d}\omega \ {\rm d}t \\
  \nonumber &=& f_0 \ u_{\rm cme} \ \tau_{\rm cme} \ \frac{\pi}{2}
  \frac{2 (\delta_{\rm cme})^2-\pi^2
    (1-\cos \delta_{\rm cme})}{(\delta_{\rm cme})^2-\pi^2},
\end{eqnarray}
with $\omega$ being the solid angle. For
\mbox{$\delta_{\rm cme} = 30\degree = \pi/6$}, this translates to
physical units as  
\begin{equation}
  \label{cme_realmass}
  M_{\rm cme,phys.} \approx
  \frac{f_0 \ u_{\rm cme} \ \tau_{\rm cme}}{16} \times 10^{14} \ {\rm kg} \ ,
\end{equation}
a typical value for a strong CME.

\subsection{CME expansion runs}

CME expansion runs have been carried out at various combinations of
CME strength, heliographic latitude, dipolar field strength, etc.
We first describe typical runs involving only a single CME, while the
case of multiple events is deferred to the ensuing section.
Unless indicated otherwise, the launch parameters $f_0=16$ and
$\tau_{\rm cme}=1$ were used.\\

\subsubsection{Single-event runs}

\begin{figure*}[t]
  \vspace*{2mm}
  \begin{center}
    \includegraphics[width=17.5cm]{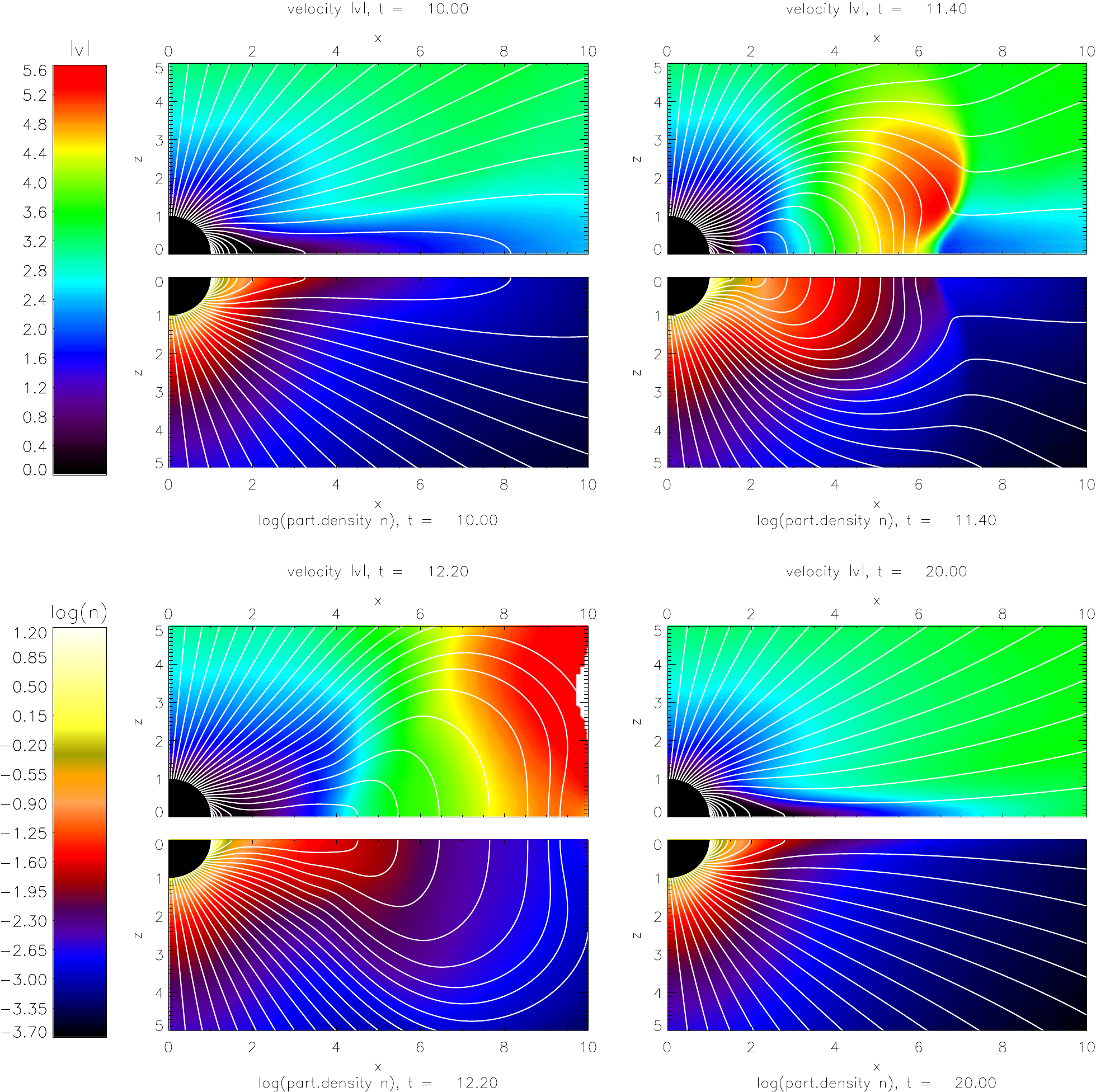}
  \end{center}
  \caption{
    \label{fig:cme-pixpanel}
    Time sequence of simulated CME expansion from pre-eruption
    ($t=10.0$) to expansion and return to equilibrium (at $t=20.0$),
    showing contours of $\|\vec{u}\|$ (top) and $\log_{10} n$ (bottom)
    in the poloidal plane ($y=0$), with magnetic field lines
    superimposed. The CME speed at onset was chosen to be
    \mbox{$u_{\rm cme}=2$}. An MPEG movie of this simulation,
      which covers the entire simulation from initialization ($t=0$)
      to convergence into steady-state (near $t=10$), CME expansion
      and back to near-equilibrium ($t\approx 20$), is available from
    the supplementary material at \supplement{mpg}.
  }
\end{figure*}

The panel of Fig.~\ref{fig:cme-pixpanel} shows selected snapshots of
a typical simulation run involving an isolated CME event. The first
frame depicts the equilibrium situation of the pre-eruptive
state. Following its initiation at the solar equator, the CME rapidly
expands outwards, thereby quickly gaining both in size and speed. Note
in particular the structures indicated by the kinked magnetic field
lines that could lead to shocks in the non-isothermal case. In the
third frame, while still continuing to accelerate, the CME reaches the
volume boundary, thereby dragging the field lines outwards and
deforming them almost radially. In the final frame, the CME has
completely left the simulation volume and the system has relaxed into
a state similar to the quiet initial situation.\\

\begin{figure*}[t]
  \vspace*{2mm}
  \begin{center}
    \includegraphics[width=17.5cm]{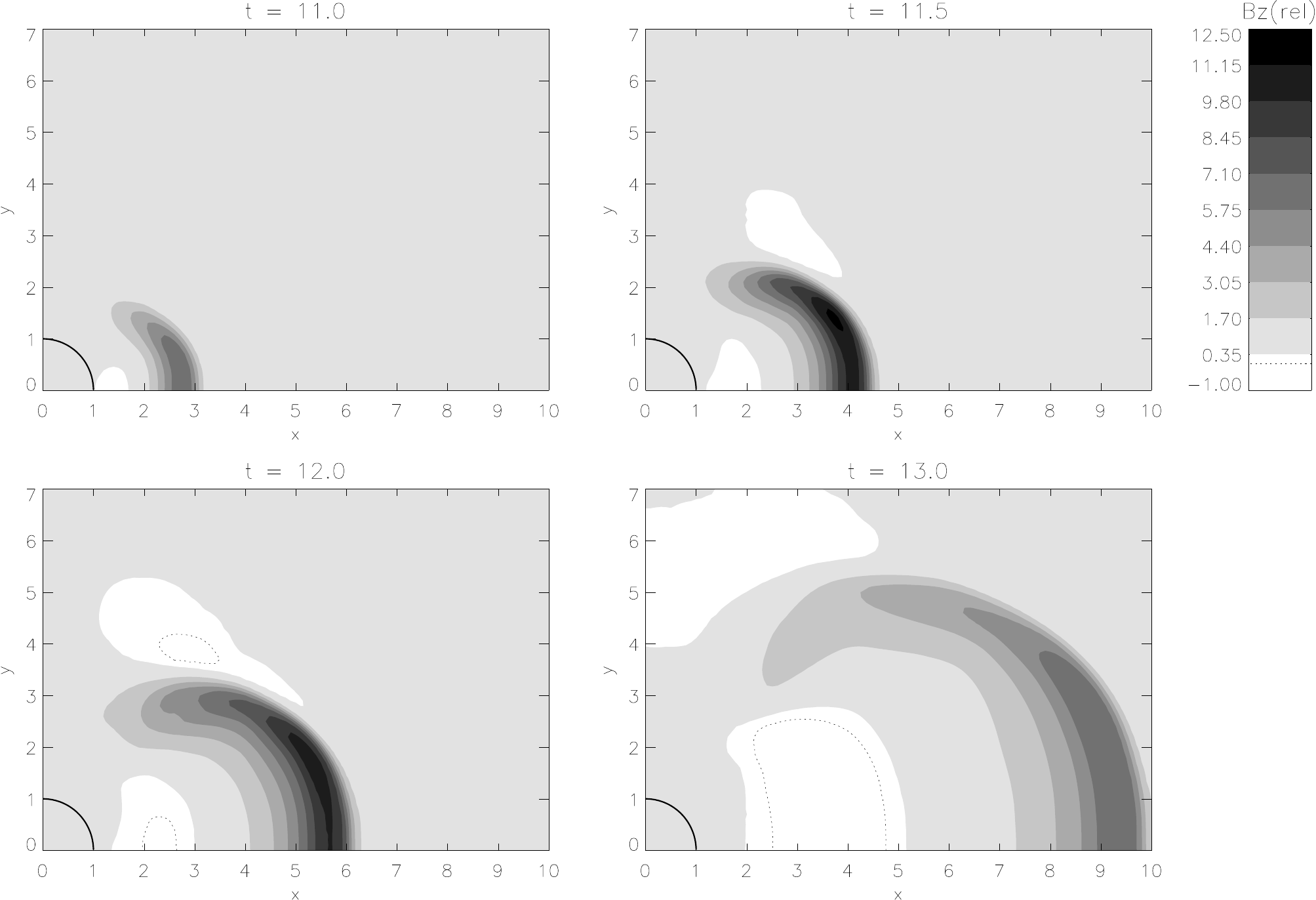}
  \end{center}
  \caption{
    \label{fig:bz-panel}
    Contour plot showing the value of $B_z$ in the $(x,y)$~plane,
    normalized to the corresponding values at the pre-eruptive
    equilibrium ($t=10.0$). The unit circle marks the position of the
    solar surface. Note the field reversals occurring within the
    encircled regions (dotted), most notably in the CME's wake.
  }
\end{figure*}

\begin{figure}
  \vspace*{2mm}
  \begin{center}
    \includegraphics[width=8.3cm]{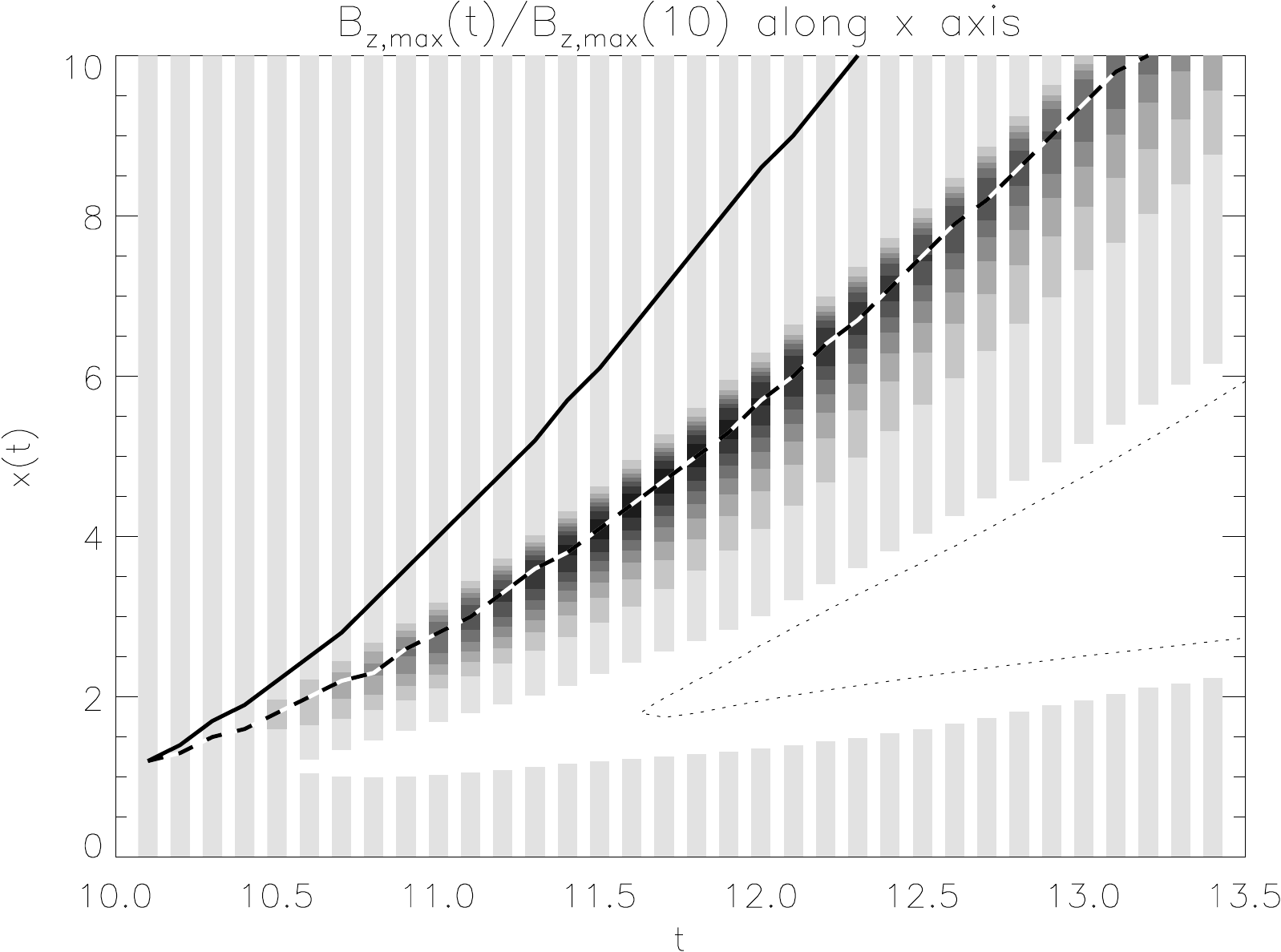}
  \end{center}
  \caption{
    \label{fig:xt-plot}
    Height-time plot tracing the position of the normalized maximum of
    $B_z$, which moves slightly ahead of the actual CME. Each vertical
    strip can be thought of as a cut along the $x$~axis of
    Fig.~\ref{fig:bz-panel} with identical color scale (including
    the dotted inversion line).
    The thick dashed line connects the respective maxima, thus
    forming an $t \mapsto x(t)$ position curve for the magnetic peak
    leading the CME. The solid line shows the corresponding $x(t)$
    plot for the faster CME ($u_{\rm cme}=2$ rather than 0). The
    corresponding contour stripes for this second CME are not shown.
  }
\end{figure}

Taking advantage of the fully 3-D nature of our simulations, we can
also access the dynamics in the perpendicular planes.
Figure~\ref{fig:bz-panel} shows time frames from the same run, this
time viewed as a contour of the sharp, wall-shaped $B_z$ signature
which arises when the CME runs into the background magnetic field and
forces it to pile up ahead of it. As can be expected, the magnetic
front moves fastest in the $x$~direction, thus forming an elongated
shell around the CME's core. On the opposite side, the CME's wake
shows a marked reduction in field strength, which even includes an
expanding region of reversed field direction trailing the CME. The
region's growing extent is particularly evident from the dotted wedge
discernible in Fig.~\ref{fig:xt-plot}. Note that the steep outward
slope of \mbox{$\| \vec{B} \|$} (\mbox{$\propto r^{-3}$} for a dipole) 
makes it necessary to normalize the values appropriately.\\

To analyses the dynamics of the CME as a whole, a reliable tracer of
its position is required.  While the CME's density shows relatively
large and irregular fluctuations which make it difficult to use it to
monitor its location, we found the magnetic field signature of
Fig.~\ref{fig:bz-panel} to be more suitable for this purpose. 
Figure~\ref{fig:xt-plot} may serve to illustrate this idea. From the
resulting $[t,x(t)]$~curves, we derive terminal velocities of
\mbox{$3.5 \ c_{\rm s} \approx 450$~km/s} and \mbox{$4.5 \ c_{\rm s}
  \approx 580$~km/s} for CMEs launched with an initial velocity of
$u_{\rm cme}=0 $ and $2$, respectively.\\

\subsubsection{Interacting CMEs}

\begin{figure*}[h!]
  \vspace*{2mm}
  \begin{center}
    \includegraphics[width=17.5cm]{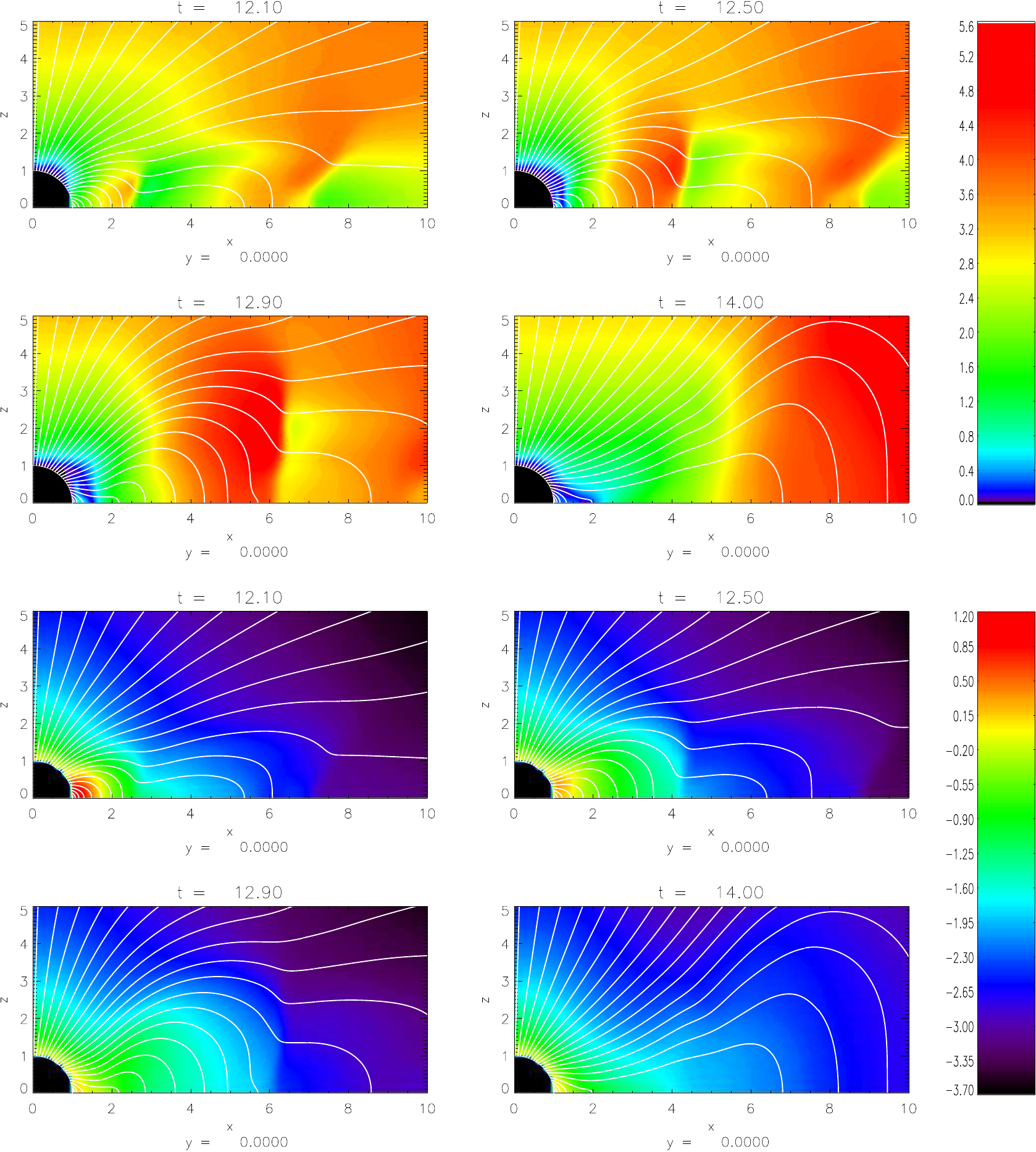}
  \end{center}
  \caption{
    \label{fig:two-cme-pixpanel}
    Selected snapshots from a simulation of two interacting CMEs
    showing velocity (top) and decadic logarithm of density (bottom),
    as well as magnetic field lines (white) in the ($y=0$) plane.
    The initial and final states are practically identical with the
    corresponding situation shown in the first and last frame of
    Fig.~\ref{fig:cme-pixpanel}, and are therefore not repeated here.
  }
\end{figure*}

With the rate of CME occurrence reaching several events per day during
solar maximum, it is not unusual to find more than one CME to be
present in a given section of interplanetary space, a fact which
motivates the numerical study of the interaction of CMEs.  
Simulations of this kind have been carried out by various authors
\citep{Vandas-etal-1997, Odstrcil-etal-2003, Schmidt-Cargill-2004,
  Wang-etal-2005}. Interacting CMEs have also been linked to the
modulation of type II radio bursts \citep{Nunes-2007}, and their
importance for SEP generation has been investigated by
\citet{Gopalswamy-etal-2005} and \citet{Vandas-Odstrcil-2004} using
2.5-D flux rope simulations. More recently, \citet{Lugaz-2008} has
connected earlier simulations \citep{Lugaz-etal-2005, Lugaz-etal-2007}
employing the BATS-R-US code \citep{Manchester-etal-2004} to actual
LASCO data by means of synthetic observations.
While it is clear that at this initial stage, our simulations cannot
be expected to rival the existing work in either detail or scientific
content, we can nevertheless demonstrate our code's general
applicability to this important sub-class of CME phenomenae. \\
Figure~\ref{fig:two-cme-pixpanel} shows selected snapshots of a
corresponding simulation run: At $t=10$, a slow ($u_{\rm cme,1}=0$)
CME is initiated along the $x$~axis, to be quickly followed by a
faster one ($u_{\rm cme,2}=2$) launched at $t=11.0$ into the same
direction. (Note that this terminology is merely used to distinguish
both entities from each other. We do not intend to relate these to the
slow/ fast dichotomy known from actual CME observations. As was shown
at the end of the preceding section, both simulated CMEs would qualify
as 'slow' in this sense.) \\ 
Both CMEs not only exhibit the individual effects of acceleration,
expansion, and field line kinking already found and discussed in the
previous case of an isolated CME, but there is apparently also a
noticeable interaction between the two as the second CME gains speed
and eventually collides with its predecessor. Note again the kinked
magnetic field lines along with a corresponding density gradient, both
related to discontinuities that would develop into shocks in the
non-isothermal case. It is also interesting to observe that the
prescribed density excess is sufficient to trigger a spontaneous,
self-consistent outward acceleration, without the need to artificially
'push' the CME forward by enforcing a non-zero initial velocity at the
instant of its launch.\\ 

\subsection{Connecting to observations}

\begin{figure}
  \vspace*{2mm}
  \begin{center}
    \includegraphics[width=8.3cm]{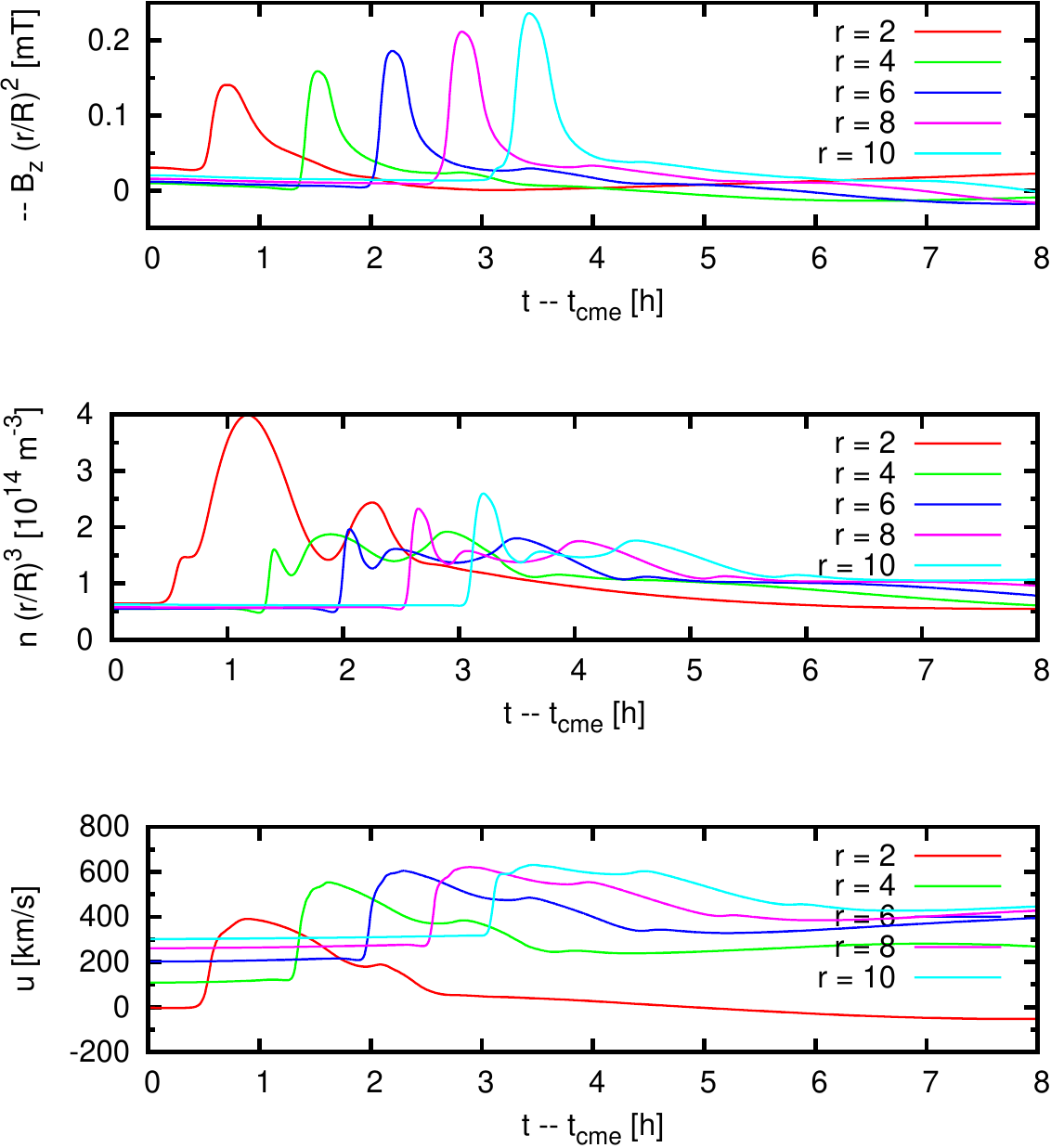}
  \end{center}
  \caption{
    \label{fig:cme-datpanel}
    Simulated profiles of magnetic field, density, and fluid velocity
    as seen by static observers situated along the CME expansion
    direction at radii \mbox{$r/R_{\odot} \in \{2,4,6,8,10 \}$} for
    the sequence of Fig.~\ref{fig:cme-pixpanel}. The minus sign at
    $B_z$ compensates for the magnetic field's north-south polarity
    (which has $B_z<0$ at $z=0$). Time is given in hours after CME
    onset. Temperature profiles are not shown due to $\gamma=1$.
  }
\end{figure}

\begin{figure}
  \vspace*{2mm}
  \begin{center}
    \includegraphics[width=8.3cm]{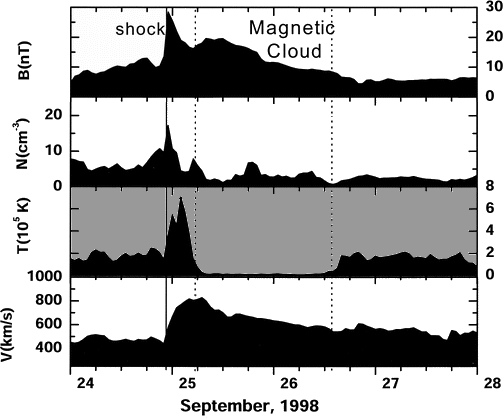} 
  \end{center}
  \caption{
    \label{fig:ace-datpanel}
    Actual in situ measurements of a magnetic cloud near 1 AU,
    adopted from \citet{Burlaga-etal-2001}. The general shape of these
    profiles is to be compared with the simulation results depicted in
    Fig.~\ref{fig:cme-datpanel}. 
  }
\end{figure}

Due to lack of in-situ data at small distances from the Sun, a direct
comparison between simulation and actual CME data is currently not
feasible. In order to at least qualitatively connect the simulations
presented here to observations, five fixed locations at radii
\mbox{$r_b \in \{2, 4, 6, 8, 10 \}$} were chosen along the CME's
trajectory (i.~e. the $x$~axis). At every time step, the values of the
non-vanishing variables \mbox{$[B_z, n, u_x]$} at these locations were
extracted from the simulation data and then combined in the panel of
Fig.~\ref{fig:cme-datpanel}. Thus, a time profile of these
quantities is generated, as it would be seen by a stationary observer
while the CME moves past his location. (It should be noted that the
profiles for particle density $n$ and magnetic field $B_z$ at radius
$r_b$ have been multiplied with $(r_b)^3$ and $-(r_b)^2$,
respectively, since otherwise the effect of radial dilution would not
have allowed curves of various radii to be presented compactly in a
single viewgraph. This obviously only changes the relative size of two
profiles against one another but leaves the shapes of individual
profiles unchanged.) \\
Using Fig.~\ref{fig:ace-datpanel}, these plots can be contrasted
with a compilation of the temporal evolution of the solar wind's MHD
properties, as measured by the Advanced Composition Explorer (ACE) for
a magnetic cloud passing the probe's location, the inner Lagrange
point at a heliocentric radius of 0.99~AU. A number of qualitative
similarities between observation and our simulation can indeed
be identified; especially the sharp rise and slow decay of the
velocity's maxima is clearly discernible in both cases. The sharp,
almost needle-shaped peaks of the magnetic field profiles also exhibit
a striking similarity. These pronounced field enhancements are induced
by the CME's compression wave, and even seem to increase further as
the driving CME accelerates outward. \\
The notable differences in the total duration of passage (about one
day for the magnetic cloud opposed to about one hour in the
simulations) can easily be accounted for by the very different sites
of observation. The cloud had much more time to extend from a
presumably rather compact object to its full length of up to
1~AU. Also, the transit time cannot be expected to be totally
independent of the duration of CME initiation (which in our case
amounts to just 1.5~h real time).
However, since observation and simulation stem from very different
heliocentric radii, a direct, quantitative comparison between the
respective profiles of Figs.~\ref{fig:cme-datpanel} and
\ref{fig:ace-datpanel} is of course not feasible. Our attempts to
identify common features between them can therefore merely serve as a
"reality check" on the general usefulness of these first simulation
runs. Besides, they may serve to illustrate the type of comparison
that are intended for future simulations covering the whole radial
range up to Earth orbit. \\

\conclusions

We have reported on the creation of a 3-D MHD model of the near-Sun
heliosphere, its numerical implementation and subsequent application
to the propagation of CMEs.\\ 
In order to adequately implement the Sun's spherical surface as an
inner boundary on the Cartesian grid, a weighted averaging procedure
was devised which is able to handle the huge gradients (most notably
of mass density) present at this boundary. The use of this procedure
also contributed to a reduction of spurious departures from the
problem's underlying symmetry, which result from the fact that the
Sun's spherical (boundary) surface cannot be mapped to a Cartesian
grid of finite cell spacing.
Comparing a Dirichlet boundary condition for the velocity against free
inward extrapolation, the latter was found to yield slightly more
accurate results, albeit requiring a more complex numerical
implementation.
To ensure a solenoidal magnetic field, the GLM scheme was found to be
inappropriate due to the presence of internal boundaries, and was thus
abandoned in favor of a classical projection method. \\ 
After the model's CWENO-based numerical realization had satisfactorily
passed various test cases, it was successfully employed to generate
stable, self-consistent MHD equilibria of the quiet, magnetized solar
wind. These were then themselves used as initial configurations to
simulate the expansion of CME-like plasma bubbles.
Since the modelling is fully three-dimensional, the CME's direction of
expansion can be chosen independently of the system's axis of
symmetry; in particular, it is possible to study expansion within the
ecliptic plane. \\
The extracted time profiles of density, flow velocity, and magnetic
field strength show qualitative similarities to actual in-situ data
obtained from satellites at much larger heliospheric distances.
The fact that such similarities can be found lends support to the
notion that the main physical processes which shape the structure of a
CME occur shortly after onset, whereas the ensuing phase of
interplanetary propagation is merely characterized by dilution and
(almost) self-similar expansion, although a direct simulation covering
the entire range up to Earth orbit will be needed to make unambiguous
statements about the CME's IP evolution and its persistent
self-similarity (or lack thereof). 
In a future extension of this work, we intend to merge heated (i.~e.\
non-isothermal) scenarios with magnetized wind models, a step which,
however desirable, could not yet be carried out due to remaining
numerical difficulties. This direction seems even more promising since
both aspects have been proven to yield satisfactory solutions
individually.\\
On the model side, we plan to include additional aspects (such as
localized heating and changes in magnetic topology) into the CME's
initialization to trigger its eruption. Since this will require a much
higher grid resolution near the solar surface, a recourse to adaptive
mesh refinement and/or parallelization becomes mandatory.

\begin{acknowledgements}
Financial support by the Wernher von Braun Foundation, the European
Commission through the SOLAIRE Network (MTRN-CT-2006-035484), and by
the DFG through the Forschergruppe 1048 (project FI 706/8-1) is
gratefully acknowledged. We also thank Ralf Kissmann for stimulating
discussions and comments, and the anonymous referees for their useful
suggestions.  
\end{acknowledgements}

\bibliography{AG2008228_ref.bib} \bibliographystyle{copernicus}

\end{document}